# Large topological Hall effect in an easy-cone ferromagnet $(Cr_{0.9}B_{0.1})Te$

Yangkun He[1,a)], Johannes Kroder[1], Jacob Gayles[1], Chenguang Fu[1], Yu Pan[1], Walter Schnelle[1], Claudia Felser[1], Gerhard H. Fecher[1]

[1]*Max Planck Institute for Chemical Physics of Solids, D-01187 Dresden, Germany*
[a)] Author to whom correspondence should be addressed: yangkun.he@cpfs.mpg.de

**Abstract**

The Berry phase understanding of electronic properties has attracted special interest in condensed matter physics, leading to phenomena such as the anomalous Hall effect and the topological Hall effect. A non-vanishing Berry phase, induced in momentum space by the band structure or in real space by a non-coplanar spin structure, is the origin of both effects. Here, we report a sign conversion of the anomalous Hall effect and a large topological Hall effect in $(Cr_{0.9}B_{0.1})Te$ single crystals. The spin reorientation from an easy-axis structure at high temperature to an easy-cone structure below 140 K leads to conversion of the Berry curvature, which influences both, anomalous and topological, Hall effects in the presence of an applied magnetic field and current. We compare and summarize the topological Hall effect in four categories with different mechanisms and have a discussion into the possible artificial fake effect of topological Hall effect in polycrystalline samples, which provides a deep understanding of the relation between spin structure and Hall properties.

In condensed matter physics, Berry phases have enabled a wider understanding of many physical concepts and phenomena, such as chiral anomalies[1,2], magnetic monopoles[3], and the anomalous Nernst effect[4]. Among them, the intrinsic anomalous Hall effect (AHE) requires the absence of time reversal symmetry and the orbital degeneracy to be lifted. The former is usually seen in ferromagnetic systems but also can be found in specific antiferromagnetic systems. The latter is due to relativistic effects such as the spin–orbit interaction, but can also be induced by a non-collinear magnetic spin texture[5,6]. The combination of these phenomena leads to momentum-space Berry curvature as a linear response to an applied electric field[7,8]. However, a real-space Berry phase originating from non-coplanar spin texture or magnetic topological excitations like skyrmions[9] with non-zero scalar spin chirality can also play the role of the magnetic field and contribute to the Hall signal, referred to as the topological Hall effect (THE)[10].

A topological Hall effect was first observed in skyrmions in non-centrosymmetric materials, such as the B20 compounds MnSi[11,12] and FeGe[13,14], in which it is stabilized by the Dzyaloshinsky–Moriya interaction. In these cubic systems, the topological Hall resistivity is usually as small as $10^{-3}$–$10^{-2}$ μΩ cm. In centrosymmetric materials with uniaxial magnetic anisotropy, such as MnNiGa,[15], the biskyrmionic phase shows a large topological Hall effect of 0.15 μΩ cm. A topological Hall effect was also observed in systems with non-coplanar antiferromagnetic spin structure, such as $Mn_5Si_3$[16], MnP[17], and $YMn_6Sn_6$[18]. Under an applied magnetic field strong enough for a metamagnetic or spin-flop transition, the (partially) antiferromagnetic coupled or canted spins align to the field direction due to the Zeeman energy, a process during which a large topological Hall resistivity of up to $10^{-1}$ μΩ cm has been observed. More recently, a topological Hall effect was also observed during the magnetization process along the hard axis of ferromagnets, such as in strong uniaxial $Cr_5Te_8$[19] and $Fe_3GeTe_2$[20] with a magnetic field applied in-plane. None of these materials, however, shows a topological Hall effect with a field along the easy axis (*c*-axis). This suggests a complex behaviour during magnetization along the hard axis in ferromagnets. Many Mn-based Heusler compounds crystallize in an inverse structure[21-25] and have a non-collinear spin structure at low temperature. They exhibit a topological Hall effect that belongs to a mixed type of the above two cases. Recently, THE was also reported in frustrated magnets[26,27]. The topological Hall effect is one of the characteristics of skyrmions, and electrical transport is easy to measure. Therefore, it can be used to select materials for potential skyrmion applications. In addition, the topological Hall effect can be used to confirm some non-coplanar spin structures without need for expensive neutron studies.

Ferromagnetism exists in a large range of compositions in $Cr_{1-x}Te$ ($0 < x < 0.4$) with different Curie temperatures $T_c$ and saturation magnetization $M_s$[28]. These compounds share a similar hexagonal structure, with Cr vacancies in every second Cr layer, while the Te layer is fully occupied. The vacancies induce small deviations from the hexagonal symmetry leading to monoclinic $Cr_3Te_4$, trigonal $Cr_2Te_3$, and trigonal or monoclinic $Cr_5Te_8$. Trigonal $Cr_5Te_8$ is a strong uniaxial ferromagnet with a magnetocrystalline anisotropy constant $K_1$ of 0.8 MJ m$^{-3}$[19,29]. However, for materials with higher Cr concentration and smaller anisotropy ($K_1 < 0.5$ MJ m$^{-3}$)[30,31], the magnetic structure is much more complicated. A canted ferromagnetic structure at low temperature was observed by neutron diffraction[32] and magnetization measurements,[28] which could lead to a possible real-space Berry phase and a topological Hall effect, providing a candidate material for skyrmion bubbles. In a previous study, we reported the magnetic structure of $(Cr_{0.9}B_{0.1})Te$[33]. Owing to the difficulty in synthesizing stoichiometric CrTe, the chromium vacancies are filled by boron, stabilizing the hexagonal structure as well as shifting the Fermi energy to modify the magnetism. The magnetic moment changes from collinear along *c* at high temperature to an easy-cone structure below the spin-reorientation transition temperature $T_{SR}$ = 140 K. The tilt angle varies with temperature.

Here we report the magneto-electronic transport properties of $(Cr_{0.9}B_{0.1})Te$ single crystals. The spin reorientation leads to a change in the Berry curvature that significantly influences both, anomalous and topological, Hall effects depending on the applied magnetic field and current direction.

Single crystals of $(Cr_{0.9}B_{0.1})Te$ were grown by an annealing process followed by water quenching. Details of the crystal growth, composition, crystal structure, magnetic properties, and electronic structure are published in Ref. [33]. The longitudinal and Hall resistivities were measured using a Quantum Design PPMS 9 with a standard four- or five-probe method.

$(Cr_{0.9}B_{0.1})Te$ crystallizes in a $B8_1$ structure (prototype: NiAs, hP4, $P6_3mmc$, 194) with alternating Cr and Te layers. The lattice constants are $a$ = 4.0184(6) Å and $c$ = 6.2684(7) Å. It is assumed that B replaces only Cr atoms of every second Cr layer. A collinear spin structure with an easy axis along *c* is observed at high temperature, whereas the magnetic moments localized at the Cr atoms become gradually tilted away from the *c*-axis at temperatures below 140 K.

The electric transport properties of $(Cr_{0.9}B_{0.1})Te$ single crystals are shown in **Fig. 1** with *H* // *c* [0001], *I* // ab plane [01-10] in Figs. 1a–c, and *H* // *ab* plane [2-1-10], *I* // *c* [0001] in Figs. 1d–f. Along both the *c*-axis and *ab* plane the longitudinal resistance shows a metallic behaviour with a kink at $T_c$ = 336 K, although the value in the *ab* plane is almost twice as large as that along the *c*-axis. The small residual-resistance ratio (RRR) is about 1.7 in-plane and 2.1 along the *c*-axis, indicating a large number of dislocations (vacancies or B atoms) inside the crystals. For magnetization along the *c*-axis, the magnetoresistance is almost zero, and as the field increases further, its value gradually decreases during heating to -1.5% at 300 K under 3 T due to spin-disorder scattering. However, the magnetoresistance is positive (1.5%) in the *ab* plane at 2 K during magnetization. With this additional effect, the negative magnetoresistance region at 3 T rises to above 200 K.

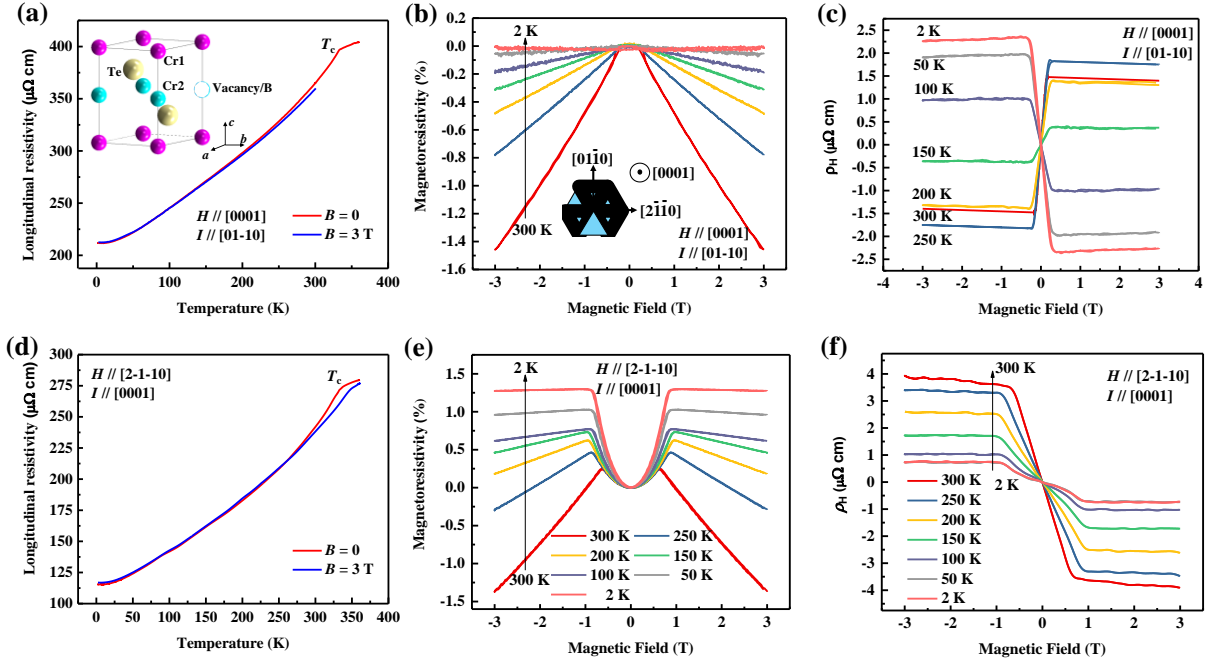

**FIG. 1.** Transport properties of $(Cr_{0.9}B_{0.1})Te$. a), d) Longitudinal resistance. b), e) Magnetoresistivity at 2–300 K. c), f) Hall signal at 2–300 K. The demagnetizing factors here are approximately 0.40 and 0.65, respectively. The crystal structure is shown as inset in a).

$(Cr_{0.9}B_{0.1})Te$ also exhibits a large, anisotropic anomalous Hall effect that depends strongly on temperature. When the applied field direction is along the $c$-axis, the anomalous Hall resistivity $\rho_{AHE}$ is positive at high temperature with collinear spin structure. However, it decreases during cooling and then changes sign at $T_{SR}$, finally reaching -2.3 μΩ cm at 2 K, as shown in Fig. 1c. When the field is in-plane, as shown in Fig. 1f, the anomalous Hall effect is always negative and changes during cooling from -3.5 μΩ cm at 300 K to about -0.7 μΩ cm at 2 K. Skew scattering is the dominant mechanism of the anomalous Hall effect for both $I // ab$ and $I // c$ [19], as shown in the linear fitting of $\rho_{AHE}$ versus longitudinal resistivity in Supplementary information. $\rho_{AHE}$ offsets the trend when the temperature approaches $T_c$. The skew-scattering mechanism confirms $(Cr_{0.9}B_{0.1})Te$ as a bad metal with a large number of defects.

The total Hall effect can be regarded as the sum of the ordinary Hall effect due to the Lorenz force, the anomalous, and the topological Hall effect using the following formula for the Hall resistivity:

$$\rho_H = R_0 B + R_s \mu_0 M + \rho_{THE}, \qquad (1)$$

where $R_0$ and $R_s$ are the ordinary and anomalous Hall coefficients, respectively. The fitted ordinary Hall resistivity is negligible, indicating a high charge carrier density of more than $10^{22}$ cm$^{-3}$; therefore, it is not shown here. A low mobility of <1 cm$^2$ V$^{-1}$ s$^{-1}$ further demonstrates that it is a bad metal, which is also confirmed by the low residual resistivity ratio (RRR) and low thermal conductivity of 3.8 W K$^{-1}$ m$^{-1}$ at 300 K. When the field is applied along the $c$-axis, $R_s$ changes sign during cooling, whereas it remains negative with the field along the $a$-axis, as shown in Supplementary information.

Moreover, the Hall signal during magnetization causes an additional effect, namely, the topological Hall effect, when $H$ // [2-1-10] and $I$ // [0001] with the easy-cone structure at low temperature, as shown in **Fig. 2**. At 250 K, with the collinear spin structure, there is no topological Hall effect. However, it appears below $T_{SR}$ of 140 K. At 2 K the value is as large as 0.21 μΩ cm. A similar result is observed when both the current and field are in two in-plane perpendicular directions. A large topological Hall effect appears near saturation, when the easy-cone structure has already been destroyed by the applied magnetic field. Note that the in-plane magnetization curve below 140 K is not linear before saturation, with a kink at around 0.2 T (Supplementary information), which is also the field at which the topological Hall effect starts to show a large value. This indicates a non-coplanar spin structure before saturation, with a solid angle $\Omega$ showing spin chirality as sketched in Fig. 3.

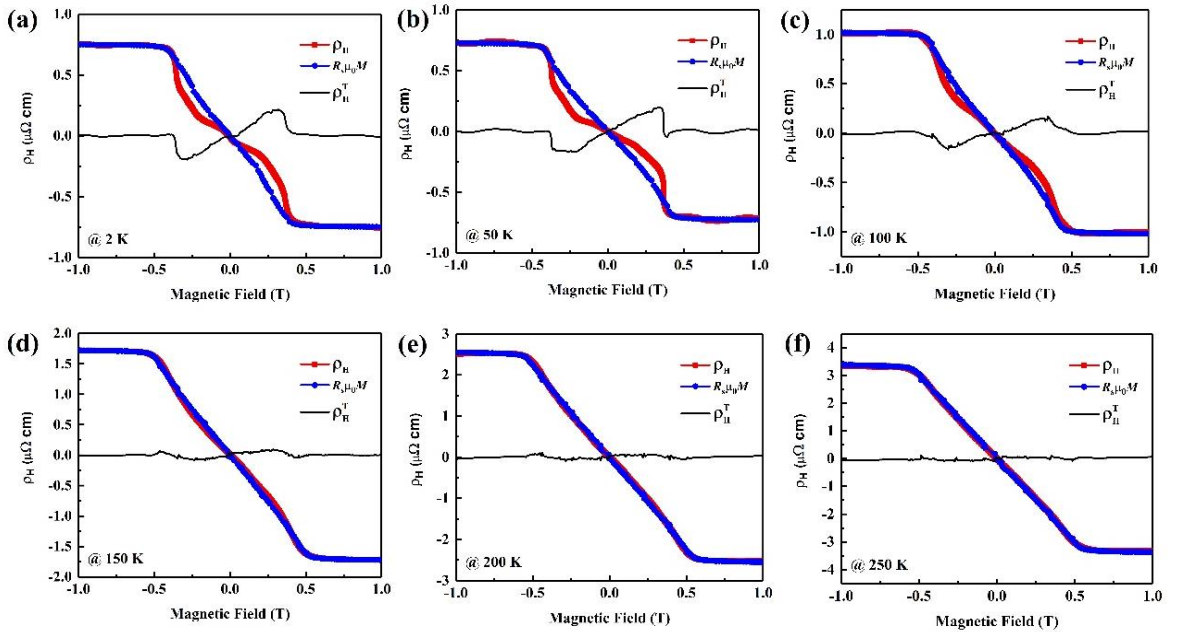

**FIG. 2.** Topological Hall effect of $(Cr_{0.9}B_{0.1})Te$ with $H \,//\, [2\text{-}1\text{-}10]$ and $I \,//\, [0001]$. The demagnetizing effect has already been corrected here to remove the possibility of an artificial effect. a)–f) correspond to different temperatures from 2 K to 250 K, respectively.

However, when the field is parallel to the *c*-axis, the topological Hall effect is too small to be distinguished from the noise. This can be explained by the lack of a non-coplanar intermediate phase during magnetization along the *c*-axis, as indicated by the absence of a kink in the magnetization curve (see Supplementary information). The domain wall motion and spin reorientation occur simultaneously during magnetization instead. This collinear spin structure does not give rise to an additional contribution to the Hall signal from the Berry curvature with $\Omega = 0$. Similar phenomena have been observed in $Cr_5Te_8$[19] and $Fe_3GeTe_2$[20], which also show a topological Hall effect when the field is along the hard axis rather than the easy *c*-axis due to the non-coplanar spin structure.

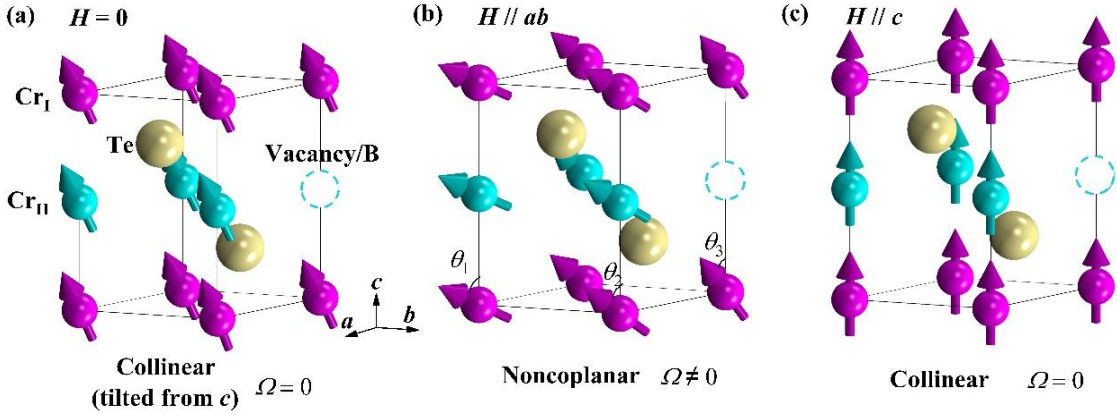

**FIG. 3.** Schematic magnetic and crystal structures of $(Cr_{0.9}B_{0.1})Te$ with a) $H = 0$, b) $H \,//\, ab$ plane, and c) $H \,//\, c$. The moment is tilted from *c* and collinear in the ground state. The in-plane field leads to a non-coplanar spin structure with different tilted angles before saturation. The moment is collinear with the *c*-axis magnetic field.

**Fig. 4** shows the phase diagram according to the above data with the field in-plane. Here $T_c$ and $T_{SR}$ are collected from the $M(T)$ curves, whereas the saturation field $\mu_0 H_s$ and the maximum field of the easy cone are collected from the $M(H)$ curves. It is clearly shown that the topological Hall effect appears at low temperature with non-coplanar spin structure near saturation, when the spin is already shifted away from the initial cone.

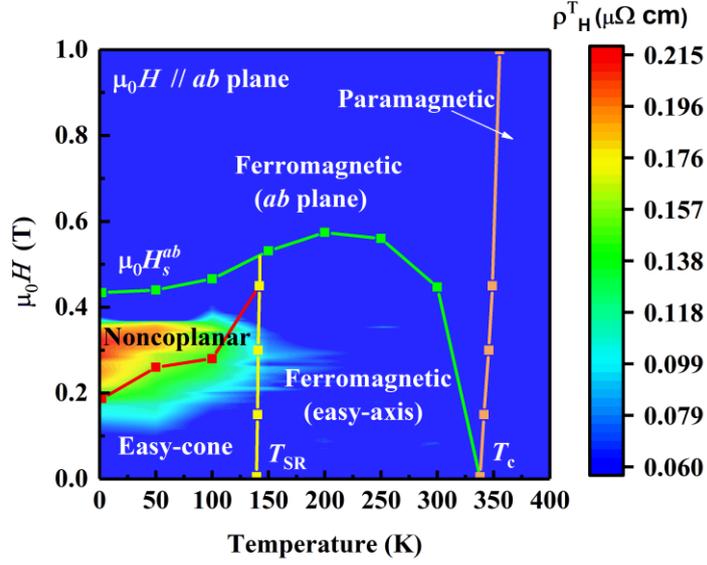

**FIG. 4.** Phase diagram of $(Cr_{0.9}B_{0.1})Te$ with in-plane field.

We compare materials exhibiting a topological Hall effect in **Table 1**. Generally, a large topological Hall effect requires a large magnetic field (see Supplementary information). The first category consists of the skyrmion materials. However, in general the topological Hall effect induced by skyrmions is small. The cubic B20 compounds, such as MnSi[11,12] or FeGe[13,14], only exhibit a topological Hall resistivity of $10^{-3}$–$10^{-2}$ $\mu\Omega$ cm. In $Mn_{1.4}PtSn$ above the spin-reorientation temperature, no topological Hall effect was found[23], although skyrmions still existed[34]. The second category covers the possibility to realize metamagnetic or spin-flop transitions under applied magnetic fields in antiferromagnetic materials with non-coplanar spin structure, as in the cases of $Mn_5Si_3$[16], MnP[17], and $YMn_6Sn_6$[18]. The third category covers magnetization from the hard axis, including $Cr_5Te_8$[19], $Fe_3GeTe_2$[20], and $(Cr_{0.9}B_{0.1})Te$. Many Mn-based Heusler compounds[21–25,35] have combined materials from all three categories. The topological Hall effects for the second and third categories are larger, with a large magnetic field, depending on the exchange coupling strength or magnetocrystalline anisotropy.

**Table 1.** Comparison of different categories of materials showing a topological Hall effect.

| Materials | MnSi, FeGe | $Mn_5Si_3$, MnP, $YMn_6Sn_6$ | $Cr_5Te_8$, $Fe_3GeTe_2$, $Cr_{0.9}B_{0.1}Te$ | $Mn_{1.4}PtSn$, $Mn_2RhSn$ |
|---|---|---|---|---|
| Ground state | Helical (AFM) | Non-collinear AFM | FM | Non-collinear FM |
| Process | Skyrmion | Metamagnetic transition | Hard-axis magnetization | Mixture |
| Field direction | All | All | Hard axis | All |
| Max THE (μΩ cm) | $10^{-3}$–$10^{-2}$ | $10^{-1}$–$10^{0}$ | $10^{-2}$–$10^{1}$ | $10^{-2}$–$10^{-1}$ |
| Magnetic field | Small | Large | Anisotropy dependent | Anisotropy dependent |

($Cr_{0.9}B_{0.1}$)Te belongs to the third group and is one of the materials that can achieve a large anomalous Hall effect with a mild field owing to its small magnetocrystalline anisotropy. Note that the topological Hall resistivity of 0.21 μΩ cm is already comparable to the anomalous Hall resistivity of 0.75 μΩ cm. Larger values of the topological Hall effect are also observed in thin films[36], which disappear when the thickness increases, indicating that the topological Hall effect is sensitive to the lattice constant, which is easily affected by strain. The small applied field is due to vanishing small single-ion anisotropy of $Cr^{3+}$ ions ($3d^3$) with a negligible orbital moment[33]. Note that the B significantly increases the Cr moment from 2.7 $\mu_B$ in the binary compound[28] to 3.1 $\mu_B$ here, and decreases the magnetocrystalline anisotropy $K_1$ from 500 kJ m$^{-3}$[31] to -100 kJ m$^{-3}$ at 2 K. This reduced field is important for potential applications.

The findings also allow us to interpret transport data from previously reported polycrystalline materials. We propose a *topological-like anomalous Hall effect* (topological AHE) that originates in polycrystals from two magnetic sublattices with different anisotropy constants, the values of which are approximately the same magnitude, but of opposite sign. We simulated a textured structure (80%*c* + 20%*a*) as an example using the data at 300 K in Fig. 1, as shown in the Supplementary information. *Note that there is no topological Hall effect at this temperature*. The magnetization along the *c*-axis saturates fast, dominating the initial magnetization curve and the anomalous Hall effect (positive). However, after saturation in the '*c*'-texture in a higher field, an additional anomalous Hall effect only emerges from the '*a*'-texture, giving a negative contribution. As result, a bump, namely, a topological-like anomalous Hall effect appears near saturation. The topological-like anomalous Hall resistivity is 1.3 μΩ cm after fitting and thus much larger than the real topological Hall resistivity observed at low temperature. Dijkstra[28] also reported Hall measurements on polycrystalline $Cr_{0.9}Te$ and $Cr_{0.8}Te$. Both samples, especially $Cr_{0.8}Te$, showed similar behaviour of two kinks before saturation, which can now be explained by topological and topological-like anomalous Hall effects. The first kink might come from the competition between anomalous Hall effects with different signs, whereas the second kink should come from the topological Hall effect of the non-coplanar spin structure. Interestingly, this topological-like anomalous Hall effect has been also realized in films in which two materials with opposite signs of the anomalous Hall effect were selected[37]. Our study points out that this can be realized in one material as well. Owing to the topological-like anomalous Hall effect great care should be taken with the transport properties of polycrystalline materials with anisotropic crystal structure to exclude the possibility of an 'artificial topological Hall effect'. For this reason, some polycrystalline materials without texture, especially those with high magnetocrystalline anisotropy, are not included in Table 1.

In conclusion, the spin reorientation transition at 140 K has a significant effect on the transport properties in $(Cr_{0.9}B_{0.1})Te$. The non-coplanar spin structure at low temperature leads to a non-vanishing Berry phase, further causing a highly anisotropic anomalous Hall effect and a topological Hall effect that strongly depends on the field direction. Consequently, a sign change of the anomalous Hall effect and a large topological Hall resistivity of 0.21 $\mu\Omega$ cm are observed. Our study provides a deep understanding of non-coplanar magnetic structure and topological Hall effect.

**Supplementary Material**

See supplementary material for the mechanism of anomalous Hall effect, topological-like anomalous Hall effect, magnetization curves and comparison of the topological Hall effect with other materials.

**Acknowledgements**

This work was financially supported by the European Research Council Advanced Grant (No. 742068) 'TOPMAT', the European Union's Horizon 2020 Research and Innovation Programme (No. 824123) 'SKYTOP', the European Union's Horizon 2020 Research and Innovation Programme (No. 766566) 'ASPIN', the Deutsche Forschungsgemeinschaft (Project-ID 258499086)

**Data availability**

Data available on request from the authors.